\documentclass[twocolumn,showpacs,preprintnumbers,amsmath,amssymb]{revtex4}

\usepackage{graphicx}
\usepackage{dcolumn}
\usepackage{bm}

\begin{document}

\preprint{prl1}

\title{Spatial synchronization and extinction of species under
external forcing }

\author{R. E. Amritkar}
\email{amritkar@prl.ernet.in}
\affiliation{%
 Physical Research Laboratory, Navrangapura,
Ahmedabad - 380009, India
}%

\author{Govindan Rangarajan}
 \email{rangaraj@math.iisc.ernet.in}
\affiliation{
Department of Mathematics,
Indian Institute of Science,  Bangalore - 560 012, India.
}%

\date{\today}

\begin{abstract}
Using a
general model we show that under a common
external forcing, the species with a quadratic saturation term
in the population dynamics first undergoes spatial
synchronization and then extinction, thereby avoiding the rescue effect. This is
because the saturation term reduces
the synchronization time scale but not the extinction time
scale. The effect can be observed even when the external forcing acts
only on some locations provided there is a synchronizing term in the dynamics.
Absence of the quadratic saturation term can help the species
to avoid extinction.
\end{abstract}


\maketitle{}

Consider two important phenomena concerning populations
of different species. First is the spatial synchronization of
populations of a species. Many examples of spatially
synchronized populations have been observed in nature
\cite{ims,post1,gren,hanski,sinclair,ranta,dwyer,sten,schwartz,lieb,blasius,post2}.
These include synchronization of vole populations by
predatory birds \cite{ims}, synchronization of caribou and
musk oxen by climate \cite{hanski,post2} and synchronization
of lynx populations probably by climate \cite{sten} and/or
dispersal \cite{schwartz}. Several other examples are
documented in Ref. \cite{lieb}. The second important phenomena
is the extinction of species. More than $99\%$ of the species
that ever existed on the surface of the earth are now extinct.
One example of such extinctions is the statistically
homogeneous $K-T$ extinction intensities observed for marine
molluscs on a global scale \cite{jablonski}. (In this letter,
extinction will mean extinction on the global scale.)
Such global extinctions are still a puzzle. It
is possible that when a species is under threat it may
survive in some isolated locations
and afterwards lead to the revival of the species. This is
the `rescue effect' \cite{allen}.
Various factors affecting extinction
such as migration, chaos, noise etc have been discussed in the
literature \cite{heino,earn}. It is believed
that spatial synchronization can promote global population
extinctions \cite{heino,earn}. However,
there is no clear understanding of the relation between spatial
synchronization and extinction and whether they will always
co-exist. This underscores the need for a general theory of
spatial synchronization and extinction of populations under
external forcing which will clarify this relation.

In this letter we investigate the time scales of spatial
synchronization and extinction. Using a general model of
population dynamics under a common external forcing we show
that the saturation term (decay term) decides these time
scales. If the dynamics has a quadratic saturation term, the
species first undergoes spatial synchronization and then
extinction. Here, extinction takes place almost
simultaneously and synchronously in all the locations.
Thus, the rescue effect \cite{allen} can not take place.
On the other had, the species which do not possess the specific
saturation term in the dynamics, will show a natural resistance
towards extinction through the rescue effect.

We present the details of our argument by
considering a general model of population dynamics. We follow
the experimental set up of Ref. \cite{ims} that considered 28
enclosed vole populations that were fenced to prevent
predatory mammals and vole dispersal. It was found that the
vole populations synchronized due to predatory birds.
Following this experimental set up, let $P_{i}(t)$ denote
the population of a species at $i-$th patch, $i =1,\ldots N$,
at time $t$ and let $Q(t)$ denote an external variable (e.g.
meteorite impacts, volcanic eruptions, predator population,
climate etc.) which interacts with the
population of the species at different patches. The coupled
dynamics can be written as
\begin{subequations}
\label{dyn_both}
\begin{eqnarray}
\frac{dP_{i}}{dt} & = & f_{1}(P_{i}(t))+
\epsilon_{1}g_{1}(P_{i}(t),Q(t))+I,
\label{dynamicsa}\\
\frac{dQ}{dt} & = & f_{2}(Q(t))+
\frac{\epsilon_{2}}{N}\sum_{i}g_{2}(P_{i}(t),Q(t)),
\label{dynamicsb}
\end{eqnarray}
\end{subequations}
where $f_{1}$  and $f_{2}$ represent the uncoupled dynamics,
$g_{1}$ and $g_{2}$ represent
the interactions and $\epsilon_{1}$ and
$\epsilon_{2}$ are interaction constants. The term $I$ represents
the interaction terms between populations of species in
different patches and is not important for our
basic argument. We will consider its effect later.

We are interested in the spatially
synchronized state, $P_{1}(t)=\ldots = P_{N}(t)= P(t)$.
Linear stability of this state can be analysed using the
following Jacobian $J$ of the vector field defined
by $P_{i}$'s and $Q$.
\begin{equation}
{\left(\begin{array}{cccc} {\frac{\partial}{\partial
P}}(f_{1}+\epsilon_{1}g_{1}) & \ldots & 0 &
\epsilon_{1}\frac{\partial g_{1}}{\partial Q}\\
\vdots & \vdots & \vdots & \vdots \\
0 & \ldots &\frac{\partial}{\partial P}(f_{1}+
\epsilon_{1}g_{1}) &
\epsilon_{1}\frac{\partial g_{1}}{\partial Q}\\
\frac{\epsilon_{2}}{N}\frac{\partial g_{2}}{\partial P} &
\ldots &
\frac{\epsilon_{2}}{N}\frac{\partial g_{2}}{\partial P} &
\frac{\partial}{\partial Q}(f_{1}+\epsilon_{1}g_{1})
\end{array}\right) }
\nonumber
\end{equation}
The eigenvectors of $J$ split into two orthogonal
subspaces $A$ and $B$ \cite{jalan}. The subspace $A$ has
dimension two and
it defines the synchronization manifold.
The subspace $B$ has dimension $N-1$  and it defines
the transverse manifold. The eigenvectors of $B$ are of
the type $\hat{\alpha}_{t}=(\alpha_{1},\ldots,
\alpha_{N}, 0)^{T}, \;
\Sigma_{i=1}^{N}\alpha_{i}=0$  and the eigenvalues are $N-1$
fold degenerate and are given by
$\frac{\partial}{\partial P}(f_{1}+\epsilon_{1}g_{1})$.
We note that under time evolution the two subspaces $A$ and
$B$ do not mix with each other and the subspace $B$ has the
same eigenvectors for all the time. Thus, it is possible to
take the time average of the degenerate eigenvalue of $B$ to
obtain the transverse Lyapunov exponent.
For the stability of the spatially synchronized state we
require the transverse Lyapunov exponent to be negative.
Imposing this, we obtain the condition for the
stability of the synchronized state as
\begin{equation}
\langle\frac{\partial}{\partial{P}}(f_{1}(P)+
\epsilon_{1}g_{1}(P,Q))\rangle<0,
\label{syncond1}
\end{equation}
where $\langle~~\rangle$ represents the time
average.

We now return to the problem of synchronization and
extinction. Near extinction, We analyse the problem by retaining the
lowest order terms in the Taylor series expansion of various
functions in Eq.~(\ref{dynamicsa}) in terms of the population
$P_{i}(t)$.
\begin{equation}
f_{1}(P_{i}) = aP_{i} - bP_{i}^{2} + {\cal{O}}(P_i^3).
\label{taylor}
\end{equation}
The first term in the expansion is a growth
term and $a>0$. The second term is
a saturation term if $b>0$. Neglecting the higher order terms,
we get the stable solution $P_{i}=b/a$. If $b<0$, then we
must include higher order terms in Eq.~(\ref{taylor}) to get
a stable solution. The interaction function $g_{1}$ to lowest
order in $P_{i}$ can be written as $-P_{i} h(Q)$
where $h(Q)$ is some function of $Q$.

Let us first consider the case $b>0$. The condition
(\ref{syncond1}) for the stability of synchronized state now
becomes
\begin{equation}
\lambda_{s}=\left\langle a- bP-\epsilon_{1}
h(Q)\right\rangle <0.
\label{syncond2}
\end{equation}
For extinction the forcing must be able to
compensate the growth and the condition for extinction
is
\begin{equation}
\lambda_{e}=\left\langle a- \epsilon_{1}h(Q)\right\rangle <0.
\label{extcond1}
\end{equation}
This condition can also be obtained by considering the
stability of $P=0$ state. Comparing the conditions (\ref{syncond2}) and
(\ref{extcond1}), we find that as
$\langle a- \epsilon_1h(Q)\rangle$ starts decreasing,
the condition
(\ref{syncond2}) will be satisfied before the extinction
condition (\ref{extcond1}) is satisfied.
If both the synchronization and extinction conditions are
satisfied then the time scale associated with
synchronization $(\tau_{s}=1/\left|\lambda_{s}\right|)$ will be
less than the time scale associated with extinction
$(\tau_{e}=1/\left|\lambda_{e}\right|)$. Thus, we conclude that
the populations in different locations will synchronize before
the extinction of the species. They will remain synchronized
as the populations at different patches start decreasing.
Hence the extinction of populations in different patches will
take place almost simultaneously.

We demonstrate that spatial synchronization precedes
extinction using a simple prey-predator model \cite{allman}.
For this model,
the different functions in Eq.~(1) are given by
$f_{1}(P)=aP-bP^{2}, g_{1}(P,Q)=-g_{2}(P,Q)=-PQ,
f_{2}(Q)=-u(Q-Q^{*})$.
We allow the predator to maintain a low equilibrium level
$Q=Q^{*}$ even when its usual prey, $P$, is rare \cite{blasius}.
For the above model, the synchronization condition becomes
$\left\langle a-2bP-\epsilon_1 Q\right\rangle<0$
and the extinction condition is given by
$\left\langle a-\epsilon_1 Q\right\rangle<0$.
In Figure~1(a), we
plot populations of different patches as a function
of time starting from random initial populations. The
parameters used are: $a=0.5,\ b=50.0,\ u=0.1,\ \epsilon_1=4.8,
\ \epsilon_{2}=1.0,\ N=100,\ Q^{*}=0.5$. We see that the populations
of different patches synchronize and then are driven to extinction.
To better understand the time scales involved,
we plot the following two parameters as a function
of time in Figure 1(b): Synchronization parameter
$S=(2/N(N-1))\sum_{i=1}^{N}\sum_{j=1}^{i-1}(P_{i}-P_{j})^2$
which measures the mean square deviation between pairs of
populations and extinction parameter
$E=(1/N)\sum_{i=1}^{N}P_{i}^{2}$
which measures the mean square populations.
We observe that initially the synchronization parameter $S$
(solid line) goes to zero with a rate greater than that of the extinction
parameter $E$ (dotted line). As the populations become very
small, $\lambda_s$ and $\lambda_e$ [Eqs. (\ref{syncond2}) and (\ref{extcond1})]
become nearly identical and the rates of decrease of $S$ and $E$
become nearly equal as can be seen from Figure 1(b).

In the dynamics of the populations we have neglected the
effect of intra-species interactions or diffusion within the
populations in different patches ($I$ in Eq.~(\ref{dynamicsa})).
It is easy to see that these effects do not affect the
conclusion of spatial synchronization before extinction. For
example, consider Eq.~(\ref{dynamicsa}) with
\begin{equation}
I = \frac{1}{N}\sum_{j=1}^{N}h_{1}(P_{j})
+\frac{1}{N-1}\sum_{j=1,j\neq i}^{N}h_{2}(P_{i},P_{j})
\end{equation}
where the first term gives a mean field type
interaction while the second term represents the interaction
between different populations, e.g. $h_{2}=P_{i}P_{j}$.
The conditions for
the stability of the synchronized state and the condition for
extinction now become, respectively,
\begin{subequations}
\begin{eqnarray}
\left\langle\left[ F(P,X,Q)\right]_{X=P}\right\rangle<0,
\label{syncond3} \\
\left\langle\left[F(P,X,Q)\right]_{X=P=0} \right\rangle<0,
\label{extcond2}
\end{eqnarray}
\end{subequations}
where $F(P,X,Q) \equiv
\frac{\partial}{\partial P}[f_{1}(P)+h_{2}(P,X)-h_{2}(X,P)
 +\epsilon_{1}g_{1}(P,Q)].$
Note that $h_{1}$ does not contribute to the above conditions
and $h_{2}$ does not contribute if it is symmetric in its
arguments, e.g. $h_{2}(X,Y)=h_{2}(Y,X)$. We see that our
conclusion about spatial synchronization before extinction is
still valid. Other effects like migration or diffusion can be
treated in a similar fashion. Again our
basic argument remains valid.

Our basic finding of synchronization before extinction can be
tested in an ecological experiment similar to those described
in Ref.~\cite{ims}. The food source of the vole population can
be decreased progressively to see whether the synchronization
persists and whether the extinction is almost simultaneous.
Similar experiments could be carried out with other
populations such as insects etc.

In the above argument we have neglected the effect of higher
order terms in the Taylor series expansion of various
functions appearing in
Eqs.~(\ref{dynamicsa}). When higher order terms are retained
two possibilities arise. First chaotic attractors can occur
and secondly there can be multiple stable solutions
\cite{strogatz}. It is believed that when the isolated patches
are individually chaotic with a weak coupling then it can lead
to asynchrony between different patches thus preventing global
extinction \cite{allen}. However, we find that such chaotic
solutions do not prevent spatial synchronization when the effect of common
external forcing is
included. It is well known in the nonlinear dynamics
literature and also as shown above, that a common forcing,
including common external noise, can synchronize chaotic
systems \cite{pikovsky,maritan}.

When there are multiple stable solutions, it is possible that
depending on the initial conditions populations in different
patches may converge to different stable solutions. In this
case extinction is difficult due to rescue effect and neither
is there a spatial synchronization.

So far we have neglected the effect of variation in parameters
from patch to patch and also the effect of local noise. Both
these factors, if they are large, can lead to spatial
asynchrony. However, we have verified that small parameter variations and
noise do not affect our conclusions. The effect of small parameter variations is demonstrated afterwards in Fig. 2.

We now discuss the case $b<0$ in Eq.~(\ref{taylor}). It is
easy to see that in this case the time scale for extinction
will be less that that for synchronization. Clearly, spatial
synchronization cannot take place before extinction. Thus, it
is is possible that the rescue effect can prevent extinction.
In Fig.~1(c), we show the poputions of different patches as
a function of time for $b=-4.0$ We see that the patches do
not show any spatial synchronization and the species may
survive due to rescue effect if the external forcing is switched
off after some time. Fig.~1(d) shows the synchronization and
extinction parameters, $S$ and $E$, as a function of time for
$b=-4.0$. We
see that initially the rate of decrease of $E$ is greater that that
of $S$ and afterwards the two rates become almost equal.

Thus, we see that the parameter $b$ can be treated as
a measure of the resistance of a species towards extinction.
Smaller the value of $b$, more is the resistance.
The second term in Eq.~(\ref{taylor}) corresponding to
parameter $b$ represents interaction between two members of a
species. The parameter $b$ is in general positive due to
competition between members. This is also reflected in various
population models used in the literature which are known to give
good fit for experimental observations
\cite{allman,murray,britton}. However, a high degree of
cooperation between the members may be able to make $b$
negative and the species more resistant to extinction.

We now consider a situation of great practical
importance. It is easy to establish
that if the coupling
parameter $\epsilon_1$ has a small variation, it does not alter the above conclusions.
However, it may happen that some patches escape the
effect of the external forcing i.e. $\epsilon_1=0$ for
these patches. We now show that our conclusions based on the
parameter $b$ are still valid provided there is some
synchronizing interaction among the patches. Let $N_2$ patches
escape the effect of external forcing and the remaining
$N_1=N-N_2$ patches be affected by the forcing. We choose
the interaction term $I = \frac{d}{N} \sum_j (P_j - P_i)$.
In this case, by using an argument similar to the one used to
show the stability of the synchronized state for Eq.~(1a), it
is possible to show that we get a two-cluster synchronized
state. The $N_1$ patches synchronize to one
value of the population say $\bar{P}_1$ and the remaining to
another value say $\bar{P}_2$. The conditions for the two
cluster synchronized state are
$\left\langle a-b\bar{P}_1-d-\epsilon_1h(Q)\right\rangle < 0$
and $\left\langle a-b\bar{P}_2-d\right\rangle < 0$ for the two
clusters respectively. The difference $\Delta \bar{P} =
\bar{P}_2 -\bar{P}_1$ evolves as
\begin{equation}
\frac{d\Delta \bar{P}}{dt}  = [a-b(\bar{P}_1 +\bar{P}_2) - d]
\Delta \bar{P} - \epsilon_{1} h(Q) \bar{P}_1
\label{dyn-diff}
\end{equation}
When the extinction condition for $N_1$ cluster is satisfied
i.e. $\left\langle a-\epsilon_{1} ch(Q) - d \right\rangle <0$ and $b>0$, the
synchronization of $\bar{P}_1$ will precede that of extinction
due to the $b$ term. The cluster $\bar{P}_2$ will also
synchronize due to $b$ term if $\bar{P}_2$ is large. Now as
$\bar{P}_1$ becomes smaller the second term on the RHS of
Eq.~(\ref{dyn-diff}) becomes small and if
$[a-b(\bar{P}_1 +\bar{P}_2) - d]<0$ then $\bar{P}_2$ will
start decreasing in some sort of generalized synchrony
with $\bar{P}_1$.

In Fig.~2 the time evolution of populations in different patches
is shown when only 50\% of the patches interact with the
external forcing. We also introduce a 5\% patch to patch
variation in all the parameters about
their respective mean values.
For $b$ positive (Fig.~2(a)) we initially see the
formation of the two-cluster synchronized state. Following this, the $N_1$ cluster
rapidly decays and is closely followed by the $N_2$
cluster. For $b$ negative (Fig.~2(b))
the populations again separate into two
distinct groups. The $N_1$ cluster shows a rapid decay but the
$N_2$ cluster shows a very small decay.
This small decay
comes from the pulling down effect of Eq.~(\ref{dyn-diff}). We note
that as the $N_1$ cluster becomes
extinct the effect of Eq.~(\ref{dyn-diff}) will also vanish and the
$N_2$ group will start its independent evolution thus escaping
extinction. 

In this letter, we have established a clear connection
between extinction and spatial synchronization of populations.
Under reasonably general conditions with external forcing we showed that spatial
synchronization precedes extinction when the parameter $b>0$
thus preventing the rescue effect. On
the other hand, for $b<0$, the species can show a natural
resistance to extinction. These conclusions are valid even if the
external forcing acts only at some locations provided there is some
synchronizing interaction between the populations.
Clearly in mass extinction events where there is a strong
common external shock, the above conclusions should be valid. Even in
other situations, we expect our general conclusions to hold
since they are based only on the parameter $b$ of the local
dynamics.

\noindent
\textbf{Acknowledgements}.
We would like to thank Professor Raghavendra Gadagkar for
helpful comments. GR was supported by a grant from ISRO, India
and is an Honorary Faculty member of the JNCASR, Bangalore,
India.


\newpage

\begin{figure}

\caption{\label{fig1} This figure demonstrates the interplay between
synchronization and
extinction in a simple prey-predator model \cite{allman}. (a) and (c)
show the populations of different patches as a function of time for
$b$ positive and negative respectively. (b) and (d) show the
synchronization and extinction parameters, $S$ and $E$, (solid and dotted lines respectively) as a
function of time.}
\end{figure}

\begin{figure}

\caption{\label{fig2} The time evolution of populations in different patches
is shown when only 50\% of the patches interact with the
external forcing.
(a) $b=50.0$; (b) $b=-4.0$.}
\end{figure}

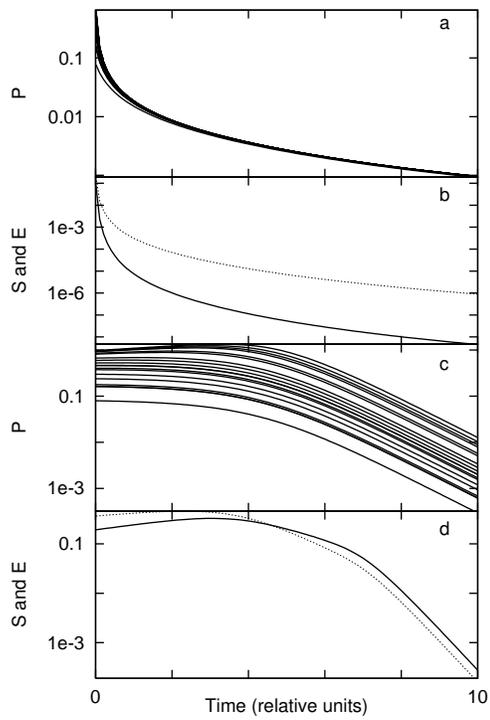
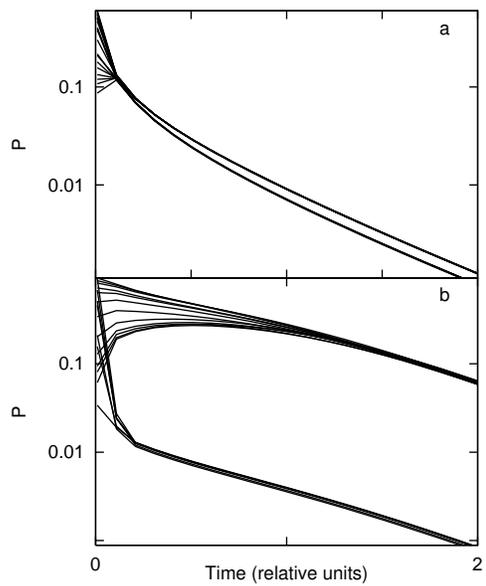
\end{document}